\documentclass[%
reprint,
superscriptaddress,
 amsmath,amssymb,
 aps,
]{revtex4-2}

\usepackage[utf8]{inputenc}
\usepackage{graphicx}
\usepackage{dcolumn}
\usepackage{bm}
\usepackage{multirow}
\usepackage[ruled,vlined]{algorithm2e}
\newcommand{\ignore}[1]{}
\usepackage{lineno}
\usepackage{color}

\usepackage{amsmath} 
\usepackage{amssymb}
\usepackage{multirow} 
\usepackage{xcolor}
\newcommand{\ket}[1]{\big|  #1 \big \rangle }

\makeatletter 
\renewcommand{\section}{\@startsection{section}{1}{0mm}
{-\baselineskip}{0.5\baselineskip}{\bf\leftline}}
\renewcommand{\subsection}{\@startsection{subsection}{1}{0mm}
{-\baselineskip}{0.5\baselineskip}{\bf\leftline}}
\makeatother

\begin{document}

\title{Authentication of quantum key distribution with post-quantum cryptography and replay attacks}

\author{Liu-Jun Wang}
\email{ljwangq@ynu.edu.cn}
\affiliation{School of Physics and Astronomy and Yunnan Key Laboratory for Quantum Information, Yunnan University, Kunming 650500, China}
\author{You-Yang Zhou}
\affiliation{School of Physics and Astronomy and Yunnan Key Laboratory for Quantum Information, Yunnan University, Kunming 650500, China}
\author{Jian-Ming Yin}
\affiliation{School of Physics and Astronomy and Yunnan Key Laboratory for Quantum Information, Yunnan University, Kunming 650500, China}
\author{Qing Chen}
\email{chenqing@ynu.edu.cn}
\affiliation{School of Physics and Astronomy and Yunnan Key Laboratory for Quantum Information, Yunnan University, Kunming 650500, China}

\begin{abstract}
With the development of quantum computers, traditional cryptographic systems are
facing more and more serious security threats. Fortunately, quantum key distribution (QKD) and post-quantum cryptography (PQC) are two cryptographic mechanisms with quantum-resistant security, and both will become important solutions for future information security. However, neither of them is perfect, and they are complementary. Quantum key distribution has unconditional security that post-quantum cryptography does not have, and PQC can provide secure and convenient authentication for QKD networks. In this paper, we propose two protocols based on PQC to realize the full authentication of the QKD data post-processing, and we only need to assume the short-term security of PQC algorithm to ensure the long-term quantum resistant security of distributed keys. We found that for the above two authentication protocols, attackers cannot successfully implement replay attacks. These authentication protocols can solve the problems of the current pre-shared key authentication in the application of large-scale quantum key distribution networks, and are expected to realize a key distribution mechanism with practical operability and quantum resistant security, which will be beneficial to promote the deployment and application of quantum key distribution networks.
\end{abstract}

\maketitle

\section*{\label{Introduction}Introduction}
Quantum computers can achieve an exponential speedup in deciphering most public key cryptography algorithms, such as RSA algorithm, discrete logarithm algorithm and Diffie-Hellman algorithm \cite{shor1994algorithms,grover1996fast,Boudot20,arute2019quantum,Zhong2021Phase}, thus posing a serious threat to encryption systems based on these algorithms. Quantum key distribution (QKD) and post-quantum Cryptography (PQC) are two known cryptographic mechanisms that are resistant to quantum computing. And through the combination of them, a more practical effective key distribution mechanism can be realized \cite{Mosca2013}.

Quantum key distribution is proposed by S. Wiesner, C. H. Bennett and G. Brassard et al. in the 1970s and 1980s \cite{wiesner1983conjugate,Bennett84}. The security of some QKD protocols has been strictly proven \cite{mayers1998quantum,lo1999unconditional,shor2000simple,Scarani09}, and more secure and efficient protocols are proposed, such as measurement-device-independent QKD protocols \cite{lo2012measurement} and twin-field QKD protocols \cite{lucamarini2018overcoming}, and these protocols have been demonstrated in experiments. At present, the point-to-point distance of key distribution has reached more than 500 km \cite{Chen2020sending,fang2020implementation,pittaluga2021600}, and the farthest distance has reached 833 km \cite{wang2022twin}. In the range of metro distance, the secure key rate can reach the order of 26 Mbps \cite{islam2017provably}. The QKD technology based on satellite platform is also developing continuously \cite{takenaka2017satellite,gunthner2017quantum,liao2017satellite,yin2020entanglement}. Although QKD has theoretical unconditional security, the practical QKD system is difficult to be made perfectly, so attacks against the QKD system have appeared from time to time \cite{lydersen2010hacking,bugge2014laser,sun2015effect,pang2020hacking}, which also indirectly prompts QKD equipment manufacturers to consider various defenses in the design to close loopholes, so as to improve the practical security of the QKD system \cite{xu2020secure}. Several quantum communication metropolitan area networks have been constructed and tested for a long period \cite{Peev09,Sasaki11,Froehlich13,wang2014field,chen2021implementation}. At the same time, using trusted relays and satellite-based QKD, it is also possible to provide quantum secure communication between cities and even continents \cite{chen2021integrated,Liao2018satellite}.

There are still some practical problems in the application of quantum key distribution, including the relatively low key rate, the difficulty of authentication, the difficulty in integrating with the existing cryptosystems, and security dependence on trusted relays. With the development of QKD protocol, optimization and technology, the key rate is gradually increasing. This paper proposes two full authentication protocols for QKD based on post-quantum cryptography, which is convenient and can ensure quantum resistant security. In addition, since post-quantum cryptography is generally based on public key cryptography, it can also provide some references for the integration of QKD and existing public key systems. At the same time, the use of authentication based on post quantum cryptography can in principle reduce the use of trusted relays in the QKD metropolitan area networks, thereby reducing the security dependence on them and improving the security of the entire quantum communication network.

For the authentication problem, this is mainly due to the unconditional security of QKD, which requires that some processes in the data post-processing must be authenticated, otherwise there will be a man-in-the-middle attack, that is, the attacker will impersonate the legitimate parties and distribute keys to each other. According to the security analysis of Ma et al. \cite{Fung2010}, the data post-processing processes requiring authentication includes basis sifting, error correction verification, random number transfer for privacy amplification, and final key verification. The authentication method currently adopted is mainly using pre-shared symmetric keys to start the first round of QKD, and then use a small part of the generated key to authenticate the subsequent round of QKD. The pre-shared key is generally realized by manually transferring the key pair. This method is secure but inconvenient to implement, especially in the QKD network, if the number of users is $n$, it is necessary to pre-share a total $C_n^2 = n(n-1)/2$ of key pairs in order to realize QKD between any two users, and when there are a large number of users, it will become very troublesome. Through trusted relays, the number of key pairs that need to be pre-shared can be reduced, but at the same time, the interconnection of the whole network is also reduced, and the security of trusted relays must be assumed.

The authentication can also be realized by the digital signature of public key cryptography, and post-quantum cryptography has quantum resistant security that traditional public key algorithms such as RSA do not have. Post-quantum cryptography mainly includes lattice ciphers, multi-variable ciphers, code-based ciphers, and hash function-based ciphers. In order to deal with the potential threat of quantum computing, in August 2016, the National Institute of Standards and Technology (NIST) launched the "Post-Quantum Cryptographic Algorithm Standardization Project" \cite{CJL+2016}, calling for PQC algorithms worldwide, and going through the third round of screening in 2020, a total of 7 algorithms were officially approved.

In the previous experiments, we verified the feasibility of applying post-quantum cryptography to QKD classical channel authentication \cite{wang2021experimental}, and conducted field experiments \cite{yang2021all}. We use the PQC signature algorithm based on lattice cipher \cite{zhang2020tweaking,Regev09,Peikert09} to authenticate the two processes of basis sifting and random number transfer for privacy amplification. However, the PQC algorithm is not used for the authentication of the error correction verification and the final key verification, but the pre-sharing key authentication is used. This is because when authenticating the above two key verification processes, the generated digest will contain part of the key information. If only the signature is made without encryption through PQC, the digest can be easily decrypted with the public key, which will reveal part of the key information. When using the pre-shared symmetric key authentication, it is equivalent to performing authentication and encryption at the same time, so there is no need to worry about information leakage. In this way, it is incomplete to authenticate only parts of QKD data post-processing through PQC, and pre-sharing pair of keys is still required. Therefore, this paper proposes to realize the complete authentication of QKD data processing based on PQC and without pre-shared keys, and ensure the quantum resistant and long-term security of QKD keys, so as to avoid the difficulties caused by authentication based on pre-shared keys, and promote the deployment and application of quantum key distribution networks.

\subsection*{Full authentication with PQC}
\begin{figure*}
\includegraphics{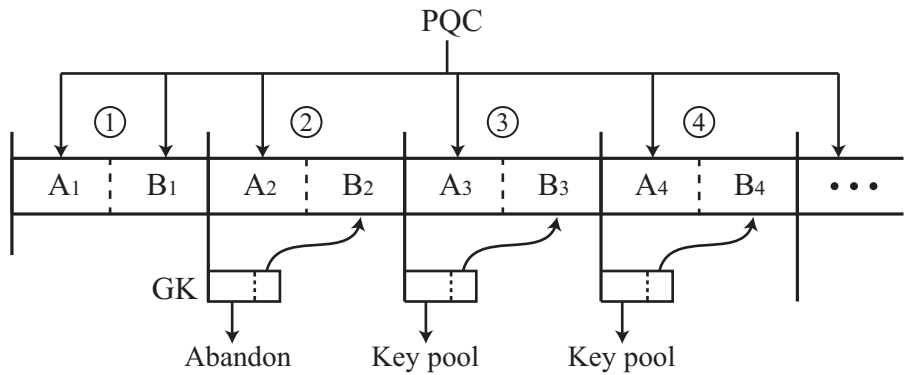}
\caption{\label{fig:sequence} \textbf{The authentication for each round of QKD and the key generation and consume.} $A_i$ and $B_i$ $(i=1, 2, 3, \cdots)$ represent the processes that require authentication for each round of QKD, where $A_i$ represents the basis sifting and the random number transfer for privacy amplification in the i-th round of QKD, which are authenticated by the PQC algorithm; $B_i$ represents the error correction verification and final key verification in the i-th round of QKD, where $B_1$ is signed and encrypted via the PQC algorithm, and $B_i$ $(i=2, 3, 4, \cdots)$ is encrypted by part of the key generated by the previous round of QKD. GK stands for generated keys.
}
\end{figure*}

To realize the full authentication of the QKD data processing by PQC and to ensure the security of the output key, it is necessary to simultaneously sign and encrypt the digests generated by the error-corrected key and the final key. The previous method was to use the pre-shared symmetric key for encryption, but since PQC does not have unconditional security, and we only want to assume its short-term security, not long-term security, Therefore, in order to ensure that PQC encryption does not affect the security of the final key, we propose to only use PQC algorithm to sign and encrypt the digest generated by the error-corrected key and the final key in the first round of QKD, and then take a part of the key generated in the first round to authenticate the second round of QKD by using symmetric key encryption method, as shown in Fig.~\ref{fig:sequence}. From the third round, each round of QKD will use a part of the key generated in the previous round to perform authentication based on symmetric key encryption. After the authentication is completed, the key for encryption is discarded, and the remaining key is stored in the key pool as secure keys.

\subsection{Data post-processing protocol 1}
Taking the BB84 protocol as an example, the sender and receiver are called Alice and Bob respectively. For each round of QKD, after the modulation, transmission and detection of the quantum signals, the first protocol of the data post-processing is as follows.

First round of QKD:

Step 1. Bob informs Alice of the positions of valid detections, and Alice discards the records of the undetected quantum states.

Step 2. Alice and Bob perform a two-way basis sifting with each other, which is authenticated by the PQC signature algorithm. If the authentication passes, continue, otherwise abort.

Step 3. Alice and Bob estimate the quantum bit error rate. If the bit error rate is higher than the threshold, the protocol will be terminated. Otherwise, the two parties will correct the raw key after the basis sifting to obtain the error-corrected key.

Step 4. Alice and Bob perform a two-way error correction verification, the digest of which is signed and encrypted through PQC algorithms. If the verification	 passes, continue, otherwise abort.

Step 5. Alice generates a string of 2n bits random numbers and sends them to Bob. The two parties negotiate to use the n bits to construct the Toeplitz matrix used for privacy amplification, and the process is authenticated by the PQC signature algorithm. If the authentication passes, continue, otherwise abort.

Step 6. Alice and Bob perform privacy amplification simultaneously to generate a secure key. And the calculation of the privacy amplification factor should take into account not only the bit error rate, but also the amount of information potentially leaked by the digest encrypted with PQC in Step 4.

Step 7. Alice and Bob perform a two-way final key verification, the digest of which is signed and encrypted through the PQC algorithms. If the verification passes, continue, otherwise abort.

Step 8. Alice and Bob construct a new Toeplitz matrix with another n bits of random numbers from Step 5. Both parties use privacy amplification to eliminate the amount of information that Eve may obtain from Step 7, and output the final key.

The data post-processing of the second round of QKD is similar to that of the first round. The difference is that Step 4 and Step 7 do not use PQC signature and encryption for authentication, but take out a part of the key generated in the first round for authentication through symmetric encryption; in Step 5 Alice only needs to generate n bits of random numbers and send them to Bob; in Step 6, the calculation of the privacy amplification factor only needs to consider the bit error rate; in Step 7, the final key is output, and there is no Step 8.

From the third round, the data post-processing of each round of QKD is basically the same as that of the second round, but in each round, Alice and Bob agree to take the same part from the final key output in the previous round for the error correction verification and final key verification in this round. After the authentication is completed, the authentication key is discarded and will not be reused.

It should be noted that the PQC signature and encryption algorithms are generally different. In the above protocol, we assume both the security of the PQC signature algorithm and the security of the PQC encryption algorithm, but this assumption is based on a short time. For example, the typical time required for the data post-processing of each round of QKD is about 1 second, then we only need to believe that the PQC algorithm is safe within 1 second. As long as the amount of keys generated by each round of QKD is greater than that required for the next round of symmetric key authentication, the secure and continuous operation of QKD can be maintained. In order to reduce the consumption of keys, considering that the basis sifting and random number transfer for privacy amplification will not leak key information, each round of QKD authentication of these two processes can be completed using the PQC signature algorithm, as shown in Fig.~\ref{fig:sequence}. If starting from the second round of QKD, the authentication of basis sifting and random number transfer does not use PQC, but uses a symmetric key for authentication, then assuming that the length of each digest is n bits (for example, SHA-256 is 256 bits), these two processes will consume n bits of keys, which will reduce the secure key rate and the maximum distance. If the duration of each round of QKD is T, the key rate will decrease
\begin{eqnarray}
\Delta R = \frac{n}{T} \ \  \text{bps}%
\label{eq:one}.
\end{eqnarray}

\begin{figure*}
\includegraphics{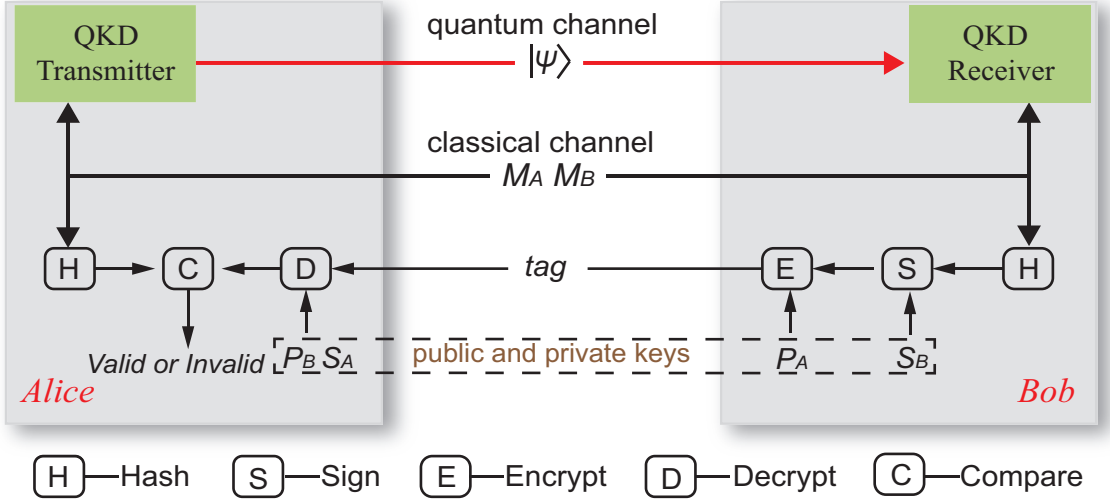}
\caption{\label{fig:Schematic} \textbf{Schematic of the signature and encryption of quantum key distribution data post-processing with post-quantum cryptography.} $M_A$ and $M_B$ represent the classic messages sent by Alice and Bob respectively, $S_A$ and $S_B$ are the private keys of Alice and Bob respectively, $P_A$ and $P_B$ are the public keys of Alice and Bob respectively, and $tag$ represents the encrypted digest. Here we only show that Alice authenticates Bob’s identity, and Bob's authentication to Alice’s identity is similar. We note that in the processes of basis sifting and random number transfer, only signatures are required, and the digest does not need to be encrypted.
}
\end{figure*}

The authentication of QKD data post-processing with PQC is shown in Fig.~\ref{fig:Schematic}, with Alice as the transmitter and Bob as the receiver. According to the post quantum cryptographic signature algorithm, each node generates a pair of public-private key pairs, such as Alice's $(S_A, P_A)$ and Bob's $(S_B, P_B)$ where $S_A, S_B$ are private keys and $P_A, P_B$ are public keys. According to the public key infrastructure protocol, the private key is kept safely by each user. The public key is handed over to the third party that everyone trusts -- the certification authority (CA), which signs it and issues it to the user in the form of a digital certificate. The CA also adopts a post-quantum signature algorithm.

At the beginning of authentication, Alice and Bob exchange digital certificates with each other, and verify the authenticity of the digital certificates with the public key of the CA, so as to obtain the public key of the other party. For the two processes of basis sifting and random number transfer required for privacy amplification, first, the QKD system generates a short digest of the message that needs to be authenticated through a hash algorithm, and passes the digest to the PQC algorithm to complete the processes of signature, encryption, transmission, decryption and comparison, as shown in Fig.~\ref{fig:Schematic}. If Alice wants to authenticate Bob's message, then Bob signs the digest with his private key $S_B$ and sends it to Alice together with the classical message. Alice decrypts the digest with Bob's public key $P_B$ and compares it with the digest generated by hashing of the received message. If they are same, the authentication passes; Otherwise, authentication fails. PQC algorithm feeds back the authentication results to the QKD system to complete this round of authentication. For the two processes of error correction verification and final key verification, it is necessary to encrypt the signed digests. According to the public key algorithm, Bob encrypts the digests with Alice's public key $P_A$ and sends the ciphertext to Alice. Obviously, the error-corrected key and final key cannot be sent. After receiving the ciphertext, Alice decrypts it with her private key $S_A$ to obtain the signed digest, and then performs the above signature verification process. Conversely, if Bob wants to authenticate Alice's message, the authentication process is similar.

\subsection{Data post-processing protocol 2}
In this protocol, we use PQC only for signatures and not for encryption. First, the PQC signature algorithm is used to authenticate the basis sifting. After the raw key is corrected, the PQC signature is used to authenticate the key consistency verification process after the correction, but the digest is not encrypted. The amount of potentially leaked information will be compressed in the subsequent privacy amplification. Once the error correction verification is valid, it indicates that Alice and Bob have identical keys. Next, the PQC signature algorithm is used to authenticate the random number transfer required for privacy amplification. After that, both parties use these random numbers to construct the same Toeplitz matrix, and perform privacy amplification on the error-corrected key. Here, it is necessary to take into account the amount of information leaked in the previous digest, so that Eve cannot grasp any information. The final key after privacy amplification is secure, but it cannot guarantee that the keys of Alice and Bob are exactly the same, so the final key verification is required, and this process needs authentication. Then Alice and Bob can take a symmetric small part of the final key according to a prior agreement for the verification of the remaining final key, so as to ensure the security of the authentication. The authentication will succeed only when the authentication keys taken out by Alice and Bob and the remaining keys to be verified are the same, otherwise, the authentication will fail. This just meets the requirements for the final key verification. Although here we uses the symmetric key encryption for authentication, it does not require a pre-shared symmetric key, nor does it need to be encrypted with PQC. Compared with the first protocol, this protocol reduces the security assumption of PQC encryption algorithm and has the same key rate.

It should be noted that although Alice and Bob have the same random basis information after basis sifting, since the basis information is not confidential, it cannot be used for authentication of the other three processes with symmetric encryption.

\subsection*{Replay attack on the authentication}
For the replay attack, it means that Eve intercepts the messages and authentication digests sent by Alice and Bob in the history, and reuses this information in a man-in-the-middle attack, trying to pretend to be Alice or Bob and establish a QKD with other parties. Whether it is basis sifting, random number transfer for privacy amplification, error correction verification, or the final key verification, the authenticated messages of these processes are all random numbers, and the message and digest of each authentication are different. Therefore, Eve must successfully execute replay attacks on all four processes. For the authentication using pre-shared keys, since the key between any two users is random, and even for the same two users, the symmetric keys used to start QKD at different times are updated, so Eve cannot use the previously intercepted encrypted digest to attack the QKD authentication between any two legitimate parties. Eve only has the possibility to launch replay attacks on authentication based on public key algorithms, including the PQC algorithm.

Basis sifting: The authentication of basis sifting is a two-way process. Here, we take Eve impersonating Alice as an example, Eve intercepts the basis sifting information and signed digest sent by Alice in history, and tries to establish a QKD link with other users. Suppose that Eve obtained string of basis $\{B_i, i=1, 2, 3, \cdots, n\}, B_i \in \{0, 1\}$, and $n$ is the length of the basis string. Taking polarization encoding as an example, $B_i=0$ represents the $Z$ basis, including horizontal polarization state $\ket H$ and vertical polarization state $\ket V$, and $B_i=1$ represents the $X$ basis, including $+45^{\circ}$ aligned state $\ket +$ and $-45^{\circ}$ aligned state $\ket -$. Note that Eve can't determine which state to send with only the basis information. In the actual QKD link, both transmission efficiency and detection efficiency are less than 1, so there must be some signals that cannot be detected by the receiver. In order to simulate the actual situation, Eve can randomly insert vacuum states between the effective signal states. At the same time, all signal states are required to be detected by the receiver to ensure that the signed digest can be replayed during the authentication of basis sifting.

Error correction verification: In this process, since only the signed digest is sent, and the error-corrected key is not sent, so Eve with a general identity cannot obtain the error-corrected key, and the quantum states corresponding to the above intercepted basis are unknown. Therefore, in order to successfully implement the replay attack in this process, Eve must have established QKD as a legitimate identity with another party in the past. For example, Bob has established QKD with Alice. At some time, Bob becomes an attacker Eve, and he tries to establish a QKD with another user Charlie via impersonating Alice. Eve has all the authentication information with Alice, including the error-corrected key and the final key, for Eve to launch a replay attack. However, as the receiver, Charlie's measurement basis is random, and the measurement results are also random. Even without considering the bit error rate, it is almost impossible to obtain the same error-corrected key required for the replay attack. The probability that the two happen to be the same is about $2^{-k}$, where $k$ is the length of the error-corrected key. Therefore, Eve cannot replay the signed digest of the error correction verification, and even the digest signed by the public key algorithm cannot be replayed. Similarly, if Alice becomes the attacker Eve at some time, she tries to establish QKD with Charlie by impersonating Bob, but at this time Charlie, as the transmitter, modulates signal states randomly, and Eve cannot replay the previous authentication digest.

Random number transfer for privacy amplification: Alice generates the random number and sends it to Bob through authentication. In this process, Bob needs to authenticate Alice’s identity. The attacker can impersonate Alice by replaying the random number and the signed digest intercepted before, so as to share the same random number with Bob and construct the same matrix for privacy amplification.

Final key verification: Since Eve cannot obtain the same corrected key, even after the same privacy amplification process, Eve cannot obtain the same final key as the intercepted key, so replay attacks cannot be implemented in this process.

Therefore, for the above three authentication methods, Eve cannot implement a complete replay attack. The main reason is that Eve cannot obtain a repetitive error-corrected key due to the randomness of the other party’s modulation states or measurement basis.

\subsection*{Conclusion and Discussion}
In this article, we propose two authentication protocols for quantum key distribution data post-processing based on post quantum cryptography. By authenticating the error correction verification and the final key verification of the first round of QKD, and encrypting the digest, and then eliminating the potentially leaked key information through privacy amplification, the long-term quantum resistant security of the final key can be realized. In the second protocol, after the authentication of basis sifting, error correction verification and random number transfer with PQC signature algorithm, Alice and Bob use a small part of the privacy amplified key to complete the authentication of the final key verification. If the authentication passes, the final key will be output. We also analyze the replay attack on QKD and obtain the fact that the attacker cannot successfully launch replay attack on the authenticated QKD.

The two authentication protocols proposed in this paper have some properties and advantages. On the one hand, they avoid the difficulty of pre-sharing symmetric keys in large-scale quantum key distribution networks. On the other hand, the protocols provide quantum resistant security. Thirdly, in combination with public key infrastructure, trusted relays are no longer required in principle within the scope of QKD metropolitan area network, so as to improve the interconnection of quantum key distribution network; Finally, we only need to assume the short-term security of the PQC algorithm, that is, after the authentication is completed, even if the PQC algorithm is cracked, the security of the generated QKD key will not be affected. This is different from the general encryption, which needs to ensure that the PQC algorithm is secure during the data confidentiality period, that is, it needs to assume the long-term security of PQC algorithm.

The authentication protocols are not only applicable to prepare and measure QKD protocols, but also to protocols such as measurement-device-independent QKD, twin-field QKD, and continuous variable QKD.

\begin{acknowledgments}
We thank T. Y. Chen, Q. Zhang and Y. Yu for valuable discussions. This work was supported by the National Natural Science Foundation of China (Grant No. 62001414, No. 12165020),  and the Yunnan Fundamental Research Project (Grant No. 202001BB050028).
\end{acknowledgments}

\bibliography{main}
\bibliographystyle{naturemag}

\end{document}